\documentclass[aps,prd,nofootinbib,preprintnumbers,superscriptaddress,twocolumn,showpacs,floatfix]{revtex4-1}
\usepackage[T1]{fontenc}
\usepackage{amsmath}
\usepackage{amssymb}
\usepackage{graphicx}
\usepackage{color}

\usepackage{hyperref}

\newcommand{\s}{\newline \vspace*{-3.5mm}}
\begin{document}
\jot = 1.2ex    

\preprint{}
\title{Light stop decays into {\boldmath $Wb\tilde{\chi}_1^0$} near the kinematic threshold}
\author{Ramona Gr\"ober}\email{groeber@roma3.infn.it}
\affiliation{INFN, Sezione di Roma Tre, 
Via della Vasca Navale 84, I-00146 Roma, Italy}
\author{Margarete M\"uhlleitner}\email{milada.muehlleitner@kit.edu}
\affiliation{Institute for Theoretical Physics, Karlsruhe Institute of 
Technology, Wolfgang-Gaede Str.~1, D-76128 Karlsruhe, Germany}
\author{Eva Popenda}\email{eva.popenda@psi.ch}
\affiliation{Paul Scherrer Institute, CH-5323 Villigen PSI, Switzerland}
\author{Alexander Wlotzka}\email{alexander.wlotzka@kit.edu}
\affiliation{Institute for Theoretical Physics, Karlsruhe Institute of 
Technology, Wolfgang-Gaede Str.~1, D-76128 Karlsruhe, Germany}
\begin{abstract}

\noindent
We investigate the decays of the light stop in
scenarios with the lightest neutralino $\tilde{\chi}_1^0$ being the
lightest supersymmetric particle, including flavour-violating (FV) 
effects. We analyse the region where the three-body decay $\tilde{t}_1
\to W b \tilde{\chi}_1^0$ is kinematically allowed and provide
a proper description of the transition region between
the three-body decay and the four-body decay $\tilde{t}_1 \to
\tilde{\chi}_1^0 b f \bar{f}'$. 
The improved treatment has been implemented in the Fortran package {\tt
  SUSY-HIT} and is used for the analysis of this region. A scan over
the parameter range including all relevant experimental constraints
reveals that the FV two-body decay into charm and
$\tilde{\chi}_1^0$ can be as important as the three-, respectively,
four-body decays if not dominant and therefore should be taken into
account in order to complete the experimental searches for the light stop.
\end{abstract}                     

\maketitle
\section{Introduction}
The discovery of a new scalar particle by the LHC experiments ATLAS
\cite{Aad:2012tfa} and CMS \cite{Chatrchyan:2012ufa} has marked a
milestone for particle physics. The immediate investigation of its
properties allowed to identify it as the Higgs boson, {\it i.e.}~the
quantum fluctuation associated with the Higgs mechanism. But still, the
question remains open if it is the Higgs boson of the Standard Model
(SM) or of some new physics (NP) extension beyond the SM (BSM). Among
the numerous NP models that are investigated, supersymmetric theories \cite{Volkov:1973ix,Wess:1974tw,Fayet:1976et,Fayet:1977yc,Fayet:1979sa,Farrar:1978xj,Dimopoulos:1981zb,Sakai:1981gr,Witten:1981nf,Nilles:1983ge,Haber:1984rc,Sohnius:1985qm,Gunion:1984yn,Gunion:1986nh,Lahanas:1986uc} certainly belong
to the best motivated and most intensely studied BSM scenarios. Based
on a symmetry between fermionic and bosonic degrees of freedom each SM
particle has a supersymmetric (SUSY) counterpart. The SUSY partners of
the top quark, the stops, play a special role. The large top quark
mass allows for a large splitting between the two stop mass eigenstates
$\tilde{t}_1$ and $\tilde{t}_2$ with interesting phenomenological
consequences. Thus, while the limits on the squarks of the first two
generations are pushed to higher and higher values
\cite{Aad:2014wea,cmssusy}, light stops have not been excluded yet by
the experiments. Stops play an important role in the corrections to the
SM-like light Higgs boson mass of the Minimal Supersymmetric Extension
of the SM (MSSM) and are crucial to shift its mass value
from the tree-level upper bound given by the $Z$ boson mass to the
experimentally measured value of $\sim 125$~GeV. Naturalness arguments 
favour the stops to be light as they significantly drive the amount
of fine-tuning at the electroweak scale \cite{Dimopoulos:1995mi}. In
the MSSM with five Higgs
bosons, two neutral CP-even ones, $h$ and $H$, one neutral CP-odd one, $A$,
and two charged scalars $H^\pm$, the maximal mixing scenario optimally
reduces the amount of fine-tuning \cite{Wymant:2012zp} while ensuring
the correct mass value of $h$. Furthermore, light stops can help for the
correct relic density through co-annihilation in scenarios with small
mass differences of 15-30~GeV between the light stop and the
lightest neutralino $\tilde{\chi}_1^0$ \cite{Boehm:1999bj,Ellis:2001nx,Balazs:2004bu,Balazs:2004ae,deSimone:2014pda,Ellis:2014ipa}. 
And last but not least, light stops
are necessary for baryogenesis to generate the matter-antimatter
asymmetry in the MSSM
\cite{Carena:1996wj,Carena:1997ki,deCarlos:1997ru,Huet:1995sh,Delepine:1996vn,Losada:1998at,Losada:1999tf,Cirigliano:2006dg,Li:2008ez,Cirigliano:2009yd,Carena:2008rt,Carena:2008vj,Laine:2012jy}. \s
 
There exist numerous experimental analyses searching for stops
in different mass windows. Light stops with masses below the
kinematic threshold for the decay into a top quark and the lightest
neutralino, assumed to be the lightest SUSY particle (LSP), can decay
through the three-body decay $\tilde{t}_1 \to W b \tilde{\chi}_1^0$
into the LSP, a $W$ boson and a bottom quark $b$. If the $\tilde{t}_1$
mass lies below the three-body decay threshold, the light stop,
assumed to be the next-to-lightest SUSY particle (NLSP), can decay
through a FV process into the LSP and a charm quark $c$
or an up quark $u$, $\tilde{t}_1 \to (u/c) \tilde{\chi}_1^0$
\cite{Hikasa:1987db,Muhlleitner:2011ww}. Another competing decay
channel in this mass regions is the 
four-body decay $\tilde{t}_1 \to \tilde{\chi}_1^0 b f \bar{f}'$
\cite{Boehm:1999tr}, where $f$ and $f'$ stand for generic light
fermions. Former bounds 
on the stop masses have been set by LEP
\cite{Abbiendi:1999yz,Abbiendi:2002mp} and Tevatron
\cite{Abazov:2008rc,Aaltonen:2012tq}. Searches based on charm tagging
and monojets have been performed by ATLAS
\cite{ctagatlas} and CMS \cite{ctagcms}. More stringent bounds have
been derived by ATLAS in decays into charm quarks or in compressed SUSY
scenarios in \cite{Aad:2014nra} as well as in final states with one
isolated lepton, jets and missing transverse momentum
\cite{Aad:2014kra}. ATLAS searches in final states with two leptons have derived
bounds on the stop mass under the assumption that it decays into a
$b$-quark and an on-shell chargino, which decays via a real or virtual
$W$ boson, or that the stop decays into a top quark and the lightest
neutralino \cite{Aad:2014qaa}. The same decay modes have been taken 
in the analysis performed by CMS \cite{Chatrchyan:2013xna}. The latter analysis
provides limits for various assumptions on the branching ratios, while
the former analyses assume branching ratios of one in the respective 
final states. \s

In \cite{Grober:2014aha} we have reinterpreted the charm-tagged and
monojet searches 
\cite{Aad:2014nra,Aad:2014kra,ctagcms} by taking into account that the branching
ratios for the FV two-body and for the four-body decay can deviate
significantly from one. This leads to considerably weakened exclusion
bounds. In this work we investigate the transition region at
the threshold of the three-body decay $\tilde{t}_1 \to W b 
\tilde{\chi}_1^0$. In particular, we analyse in this threshold
region the interplay between the FV two-body decay and the three-body
decay above the threshold, respectively, the four-body decay just below the
threshold.\footnote{Note that we choose the parameters such that in the
four-body decay only the diagrams with the intermediate $W$ boson can
become on-shell in the investigated region.}
It turns out that the two-body decay can still be
significant here for certain parameter configurations and can hence be
exploited to improve and/or complement analyses based on the
three-body decay final states. We extend and refine former analyses
\cite{Porod:1996at,Porod:1998yp,Djouadi:2000bx} by including the
recently computed SUSY-QCD corrections to the FV two-body decay
\cite{Grober:2014aha}\footnote{See also \cite{Aebischer:2014lfa}.} and
the FV tree-level couplings in the three-body decay as well as in the
four-body decay where also a non-vanishing $\tau$ and bottom mass in
the final state \cite{Grober:2014aha} are taken into account. 
Furthermore, the transition region between three- and 
four-body decays is consistently described by including a finite
width in the $W$ boson propagator, which becomes virtual below the
three-body decay threshold. Finally, we check for the accordance with the LHC data
on the Higgs boson search, the exclusion limits from SUSY searches as
well as constraints from the relic density and $B$-physics and from electroweak
precision measurements. \s

In Sec.~\ref{Wbchi} details on the calculation of the three- and four-body
decay widths are given, followed in Sec.~\ref{numerical} by the
description of the numerical analysis and the applied
constraints. Our results are presented in Sec.~\ref{results}. We
conclude in Sec.~\ref{conclusion}.  

\section{Three- and Four-Body Stop Decays \label{Wbchi}}
\begin{figure*}
 \includegraphics[width=14cm]{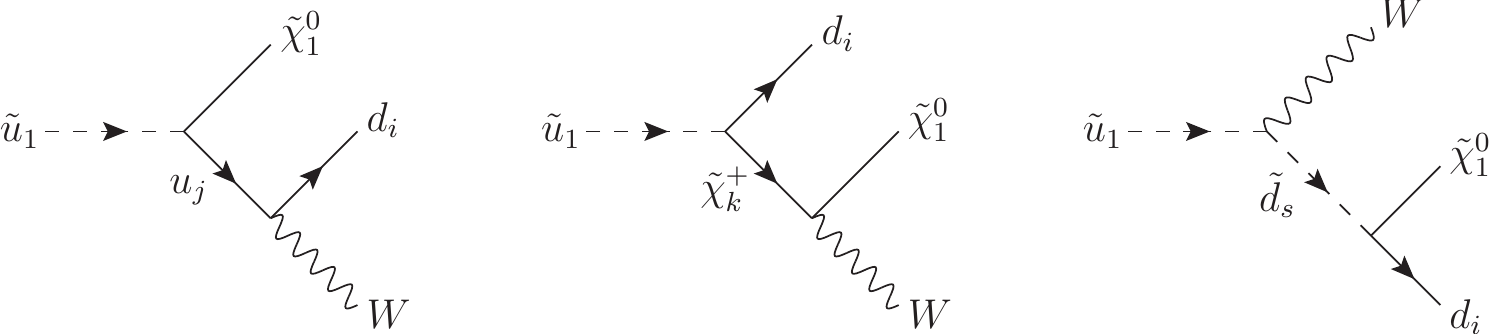}
\caption{Feynman diagrams of the process $\tilde{u}_1\to W d_i
  \tilde{\chi}^0_1$ ($i,j=1,2,3, s=1,...,6, k=1,2$). \label{fig:feyndia3bod}}
\end{figure*}

We work in the framework of the MSSM with general flavour
structure. Flavour-violating effects are constrained by experiment to
be very small which can be naturally accounted for in  the Minimal
Flavour Violation (MFV) \cite{Chivukula:1987fw,Hall:1990ac,Buras:2000dm,D'Ambrosio:2002ex,Bobeth:2005ck} approach {\it e.g.}, where the only sources of 
FV are given by the CKM matrix elements. Flavour violation is induced
through renormalisation group running. Due to the large mixing in the
stop sector, the lightest up-type squark $\tilde{u}_1$ is then mostly
stop-like. For convenience, we occasionally refer to it as the light stop
$\tilde{t}_1$ in the following although it is understood that it has 
small flavour admixtures from the charm- and up-flavours.
The three-body decay of $\tilde{u}_1$ into the lightest
neutralino, a down-type fermion $d_i$ ($i=1,2,3$), where $i$ denotes
the quark flavour, and a $W$ boson, 
\begin{equation}
\tilde{u}_1 \to W d_i \tilde{\chi}^0_1 \;,
\end{equation}
proceeds via down-type squark, chargino and up-type quark exchange. 
The Feynman diagrams are displayed in Fig.~\ref{fig:feyndia3bod}. The
index $s=1,...,6$ of the exchanged squark refers to one of the six squark mass 
eigenstates, which are not flavour eigenstates any more. In case of
small FV as given in the MFV setup, the dominant final
state is given by $W b \tilde{\chi}^0_1$. We have 
calculated the three-body decay with the general flavour 
structure by extending the results of~\cite{Porod:1996at,
  Djouadi:2000bx} to all flavours. The full dependence on the
bottom quark mass has been taken into account, whereas the first and
second generation quark masses have been set to zero. The result for
the decay width has been checked against a second, independent calculation
by using {\tt FeynArts/FormCalc}
\cite{Hahn:2000kx,Hahn:1998yk,Hahn:2006qw,Hahn:2006zy}. \s

In the threshold region where the three-body decay mode of the light stop
into $Wb\tilde{\chi}^0_1$ opens up,
the off-shell effects of the $W$ boson can be described by
implementing the $W$ boson width in the propagators of the 
$W$ boson diagrams in the four-body decays 
\begin{eqnarray}
\tilde{u}_1\to \tilde{\chi}_1^0 d_i f\bar{f}' \;.
\end{eqnarray}
Again in case of small FV the dominant final state is the one
involving the $b$-quark, {\it i.e.} $d_i = b$. 
The $W$ boson width in the propagators introduces a gauge
dependence. The width renders the $W$ boson mass
$m_W$ in the $W$ boson propagators complex, whereas it is real in the
corresponding Goldstone boson couplings, so that the cancellation of
the gauge parameter dependence between the $W$ boson and the
associated Goldstone boson diagrams cannot take place any
more. Possible solutions are given by the complex mass scheme
\cite{Denner:1999gp}, where a complex mass is introduced 
also in the Feynman rules, or by the overall-factor
scheme~\cite{Baur:1991pp, Baur:1995aa}, in which the whole tree-level
amplitude is multiplied by
\begin{equation}
\prod_{W \text{propagators}}\frac{p_W^2-m_W^2}{p_W^2-m_W^2+i m_W \Gamma_W}\;,
\end{equation}
where $p_W$ denotes the $W$ boson four momentum and $\Gamma_W$
the $W$ boson width. The product $\prod$ accounts for the maximal number of
$W$ propagators in the amplitude. We use the overall-factor scheme to
ensure a gauge independent result. The drawback of this method is that
close to the threshold the non-resonant contributions are neglected. 
We checked, however, explicitly, that in the scenarios found in our 
numerical analysis below, the effect of neglecting the non-resonant
contribution is less than about 2\% and hence acceptable. The three-body decay
and the thus calculated four-body decay widths have been implemented in the {\tt
  SDECAY}~\cite{Muhlleitner:2003vg, Muhlleitner:2004mka} routine of 
{\tt SUSY-HIT}~\cite{Djouadi:2006bz}, where the SUSY Les Houches
Accord (SLHA) \cite{Skands:2003cj} format has been extended to the SLHA2
format~\cite{Allanach:2008qq}, as described in
Ref.~\cite{Grober:2014aha}, to account for FV. \s

In order to ensure that the three-body and the four-body decay
widths match for $m_{\tilde{u}_1}-m_{\tilde{\chi}_1^0}$ mass differences above
the kinematic threshold of an on-shell $W$ boson, the $W$ boson width must
be computed in accordance with the loop order and the input values used for the
computation of the four-body decay width. Thus, the tree-level $W$
boson decay width is computed with massless first and second
generation fermions, while the masses of the bottom quark and the
$\tau$ lepton are kept finite.

\section{Numerical Setup and Experimental
  Constraints \label{numerical}}
We have performed a random scan over the parameter space of the model
with the same settings as in the $U(2)$-inspired scan of
Ref.~\cite{Grober:2014aha}. The parameters have been varied in the ranges
\begin{align}
 \tan \beta & \in [1,15] \ , \nonumber\\
 M_A & \in [150,1000]\ \text{GeV} \ , \nonumber\\
 M_1 & \in [75,500]\ \text{GeV} \ , \nonumber\\
 M_{\tilde{U}_3} & \in [300,600]\ \text{GeV} \ , \nonumber\\
 M_{\tilde{Q}_3} & \in [1000,1500]\ \text{GeV} \ , \nonumber\\
 A_t & \in [1000,2000]\ \text{GeV}\ .
 \end{align}
The remaining MSSM input parameters have been chosen as
\begin{align}
 M_2 & = 650 \ \text{GeV} \ , \nonumber \\
 M_3 & = 1530 \ \text{GeV} \ , \nonumber \\
 \mu & = 900 \ \text{GeV} \ , \nonumber \\
 M_{\tilde{E}_{1,2,3}} & = M_{\tilde{L}_{1,2,3}} = 1 \ \text{TeV} \ ,
 \nonumber \\
 M_{\tilde{Q}_{1,2}} & = M_{\tilde{U}_{1,2}}  = M_{\tilde{D}_{1,2,3}}
 = 1.5 \ \text{TeV} \ , \nonumber \\
 A_U & = A_E = A_D  = 0 \ .
\end{align}
The SM input parameters have been set to the PDG values \cite{pdg2014}. 
We have applied the same constraints on the
generated parameter points as in \cite{Grober:2014aha}, but updated
the branching ratio of the $B_s^0 \rightarrow \mu^+ \mu^-$ decay to the
recently reported value 
\begin{equation}
 \mathcal{B}(B_s^0 \rightarrow \mu^+ \mu^-) = \left( 2.8^{+0.7}_{-0.6} \right) \times 10^{-9} \quad \text{\cite{CMS:2014xfa}}\ .
\end{equation}
Additionally we have checked for the dominant restrictions due to
electroweak precision observables by throwing away all points which
are outside the 2$\sigma$ interval around the  experimental value for
the $\rho$-parameter  
\begin{equation}
 \rho = 1.0004 \pm 0.00024 \quad \text{\cite{pdg2014}} \;.
\end{equation}
Among the parameter points fulfilling the constraints we have retained
only those, for which the masses of the
lightest up-type squark $\tilde{u}_1$ and the lightest neutralino
$\tilde{\chi}_1^0$ comply with 
\begin{equation}
m_{\tilde{u}_1}-m_{\tilde{\chi}_1^0} \in [60,\ 140]\ \mathrm{GeV} \ . \label{eq:numerics1}
\end{equation}
The mass window around the three- to four-body decay threshold has
been chosen large enough to allow for studying all effects
that emerge in the threshold region. 
Finally, for the parameter points above the threshold {\tt{SModelS}}
\cite{smodels:v1,Kraml:2013mwa,smodels:wiki} based on
the tools {\tt Phythia}~6.4 \cite{Sjostrand:2006za}, {\tt NLL-fast}
\cite{Beenakker:1996ch,Beenakker:1997ut,Kulesza:2008jb,Kulesza:2009kq,Beenakker:2009ha,Beenakker:2010nq,Beenakker:2011fu}
and {\tt PySLHA} \cite{Buckley:2013jua}, 
is used to ensure that all parameter points fulfil the exclusion
bounds derived from direct searches by ATLAS and CMS
\cite{Aad:2014qaa,Aad:2014kra,Aad:2014nra,Chatrchyan:2013xna,ctagcms,cmsrazor}. Since
the searches in the FV two-body decay channel are not
covered by {\tt{SModelS}} yet, for the parameter points below the
threshold the procedure explained in \cite{Grober:2014aha} is
used. The scenarios surviving all constraints include chargino masses
around 660~GeV, slepton masses of ${\cal O}(1~\mbox{ TeV})$ and
charged Higgs masses in the range $\sim 400$ to $\sim 1$~TeV, so that
the corresponding diagrams in the four-body decay with these particles
in the intermediate propagators never go on-shell in the investigated
threshold region. 

\section{Results\label{results}}
\begin{figure}[t!]
\begin{center}
\includegraphics[scale=0.4]{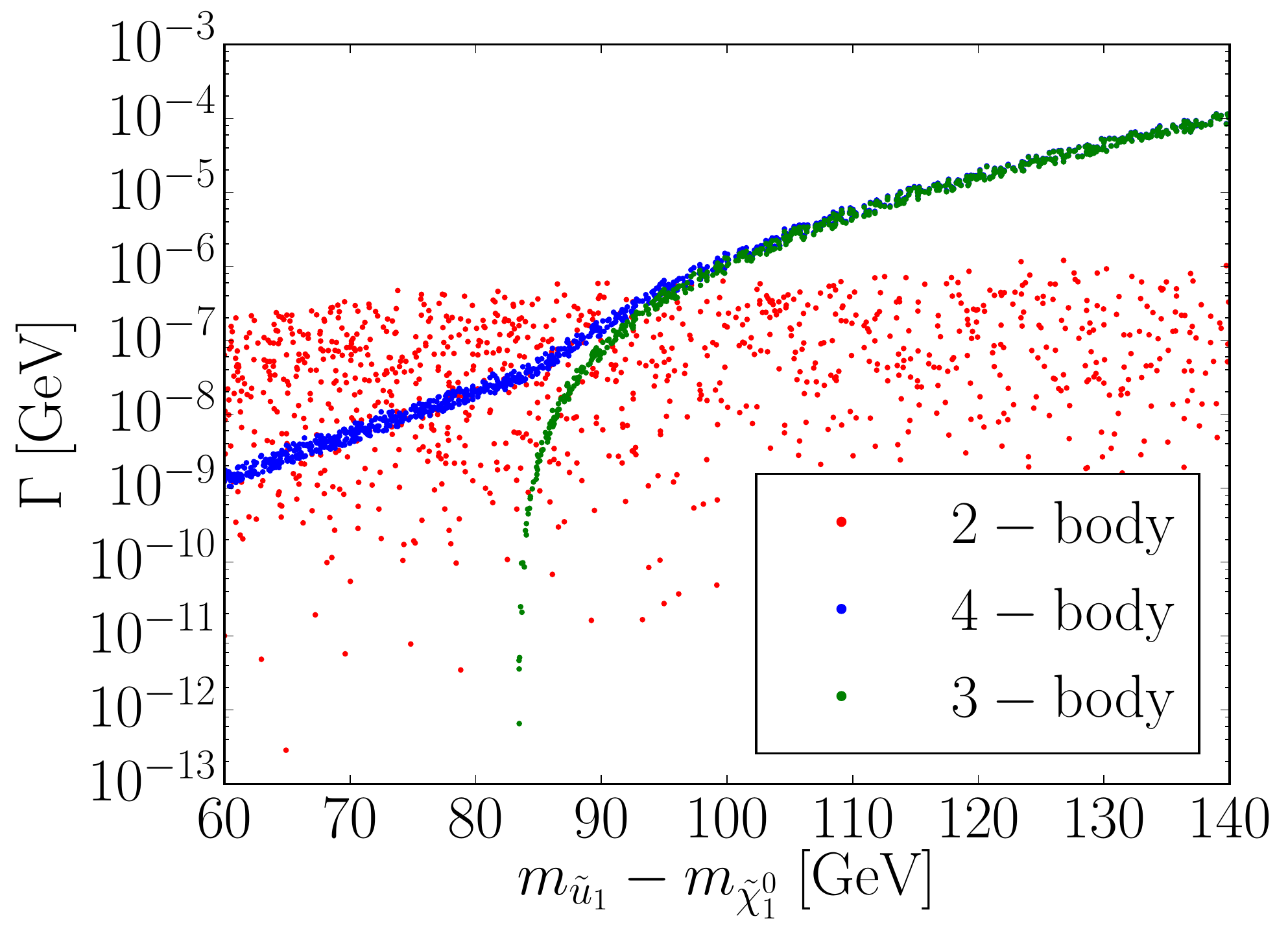}
\end{center}
\vspace{-6mm}\caption{Partial widths of the $\tilde{u}_1$ two- (red),
  three- (green) and four-body (blue) decays as a function of the
$m_{\tilde{u}_1}-m_{\tilde{\chi}_1^0}$ mass difference.\label{fig:widths}} 
\end{figure}

\begin{figure}[b!]
\begin{center}
\includegraphics[scale=0.4]{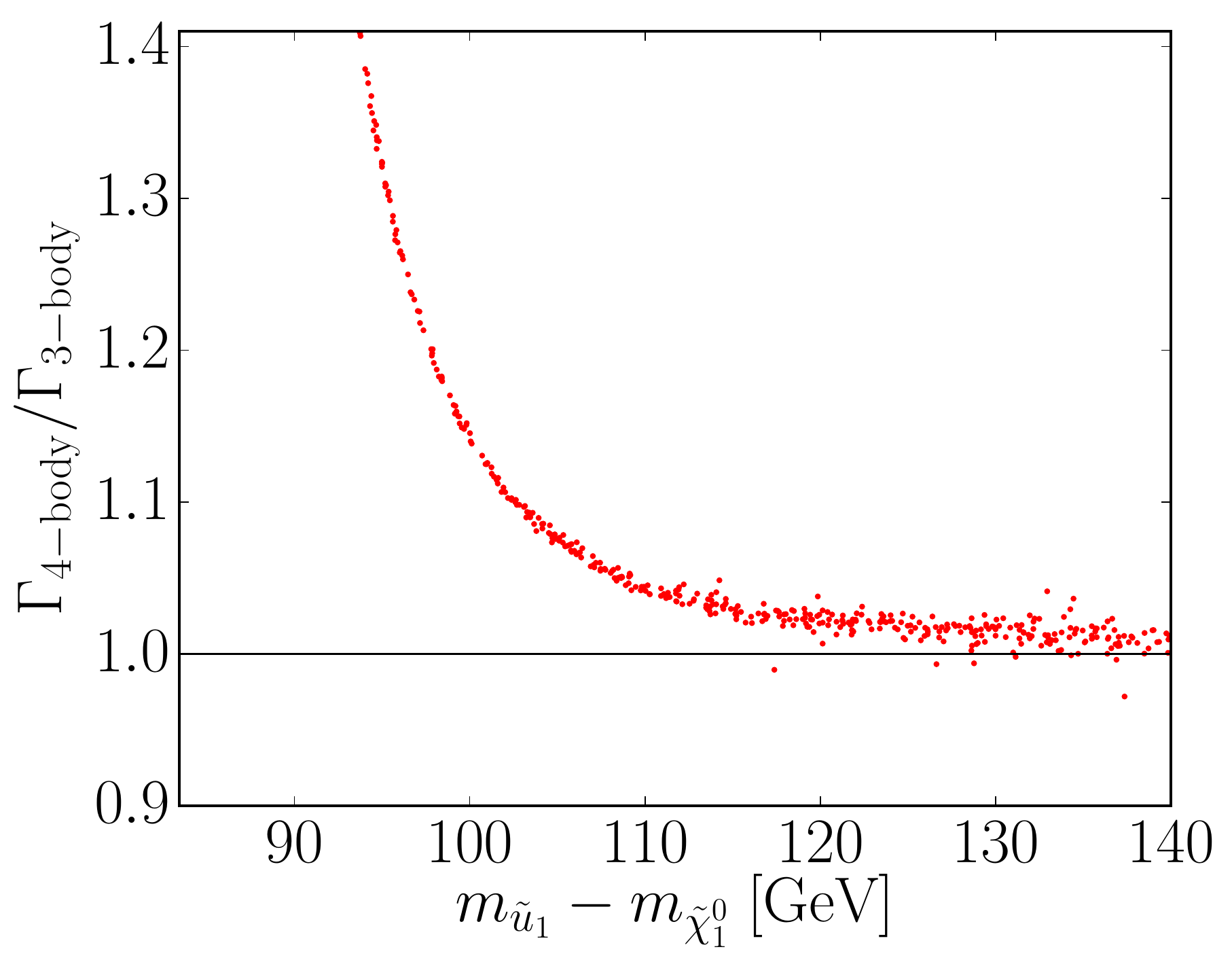}
\end{center}
\vspace{-6mm}\caption{Comparison of the $\tilde{u}_1$ four- and
  three-body decay widths as a function of the
  $m_{\tilde{u}_1}-m_{\tilde{\chi}_1^0}$ mass difference. \label{fig:compfourthree}}
\end{figure}

Figure~\ref{fig:widths} shows the two-, three-, and four-body decay
widths, respectively, for the parameter points of our scan which are
in accordance with all applied constraints. 
The three-body decay, given by the green points, sets in at the the
threshold $m_{\tilde{u}_1}-m_{\tilde{\chi}_1^0} = m_W +m_{d_i}$. As
expected, it approaches the four-body decay, illustrated by the blue points, for
$m_{\tilde{u}_1}-m_{\tilde{\chi}_1^0}$ mass differences sufficiently above the
threshold.\footnote{Note, that we have implemented the total width of the top quark in the three-body decay and explicitly checked that the top width does not play a role in the three-body decay, also for $m_{\tilde{u}_1} - m_{\tilde{\chi}_1^0}$ mass differences as large as 140~GeV. The three-body decay with the top quark width and FV couplings has been implemented in the SUSY-HIT version that includes FV decays.} The relative size of the four- and the three-body decay widths is
displayed in Fig.~\ref{fig:compfourthree}. It shows that the finite width
effects are still sizeable $20\ \text{GeV}$ above the threshold and
therefore should be taken into account, as is done by including the 
total width of the $W$ boson in the four-body decay. Note, that the
scattering of the points at the upper end of the mass difference is 
\begin{figure}[t!]
\begin{center}
\includegraphics[scale=0.4]{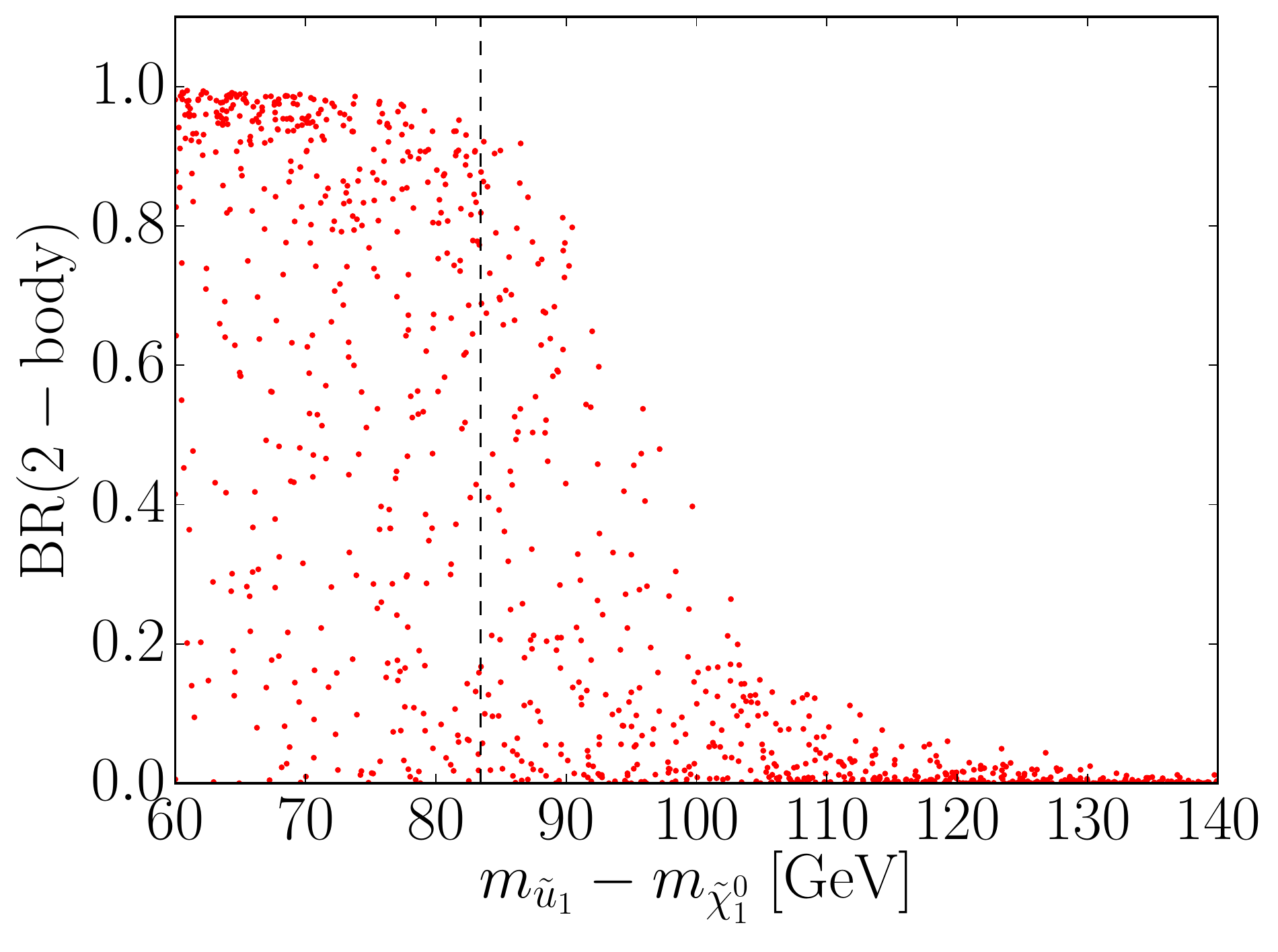}
\end{center}
\vspace{-6mm}\caption{Branching ratio of the $\tilde{u}_1$ two-body
  decay as a function of the $m_{\tilde{u}_1}-m_{\tilde{\chi}_1^0}$
  mass difference. The dashed 
  line marks the threshold for the three-body decay. \label{fig:2bodybran}}
\end{figure}
subject to the
numerical integration precision in the four-body decay. Furthermore,
the remaining off-set between the four- and three-body decay at large mass
differences is due to the finite value of the $W$ boson width. \s
\begin{figure}[b!]
\begin{center}
\includegraphics[scale=0.4]{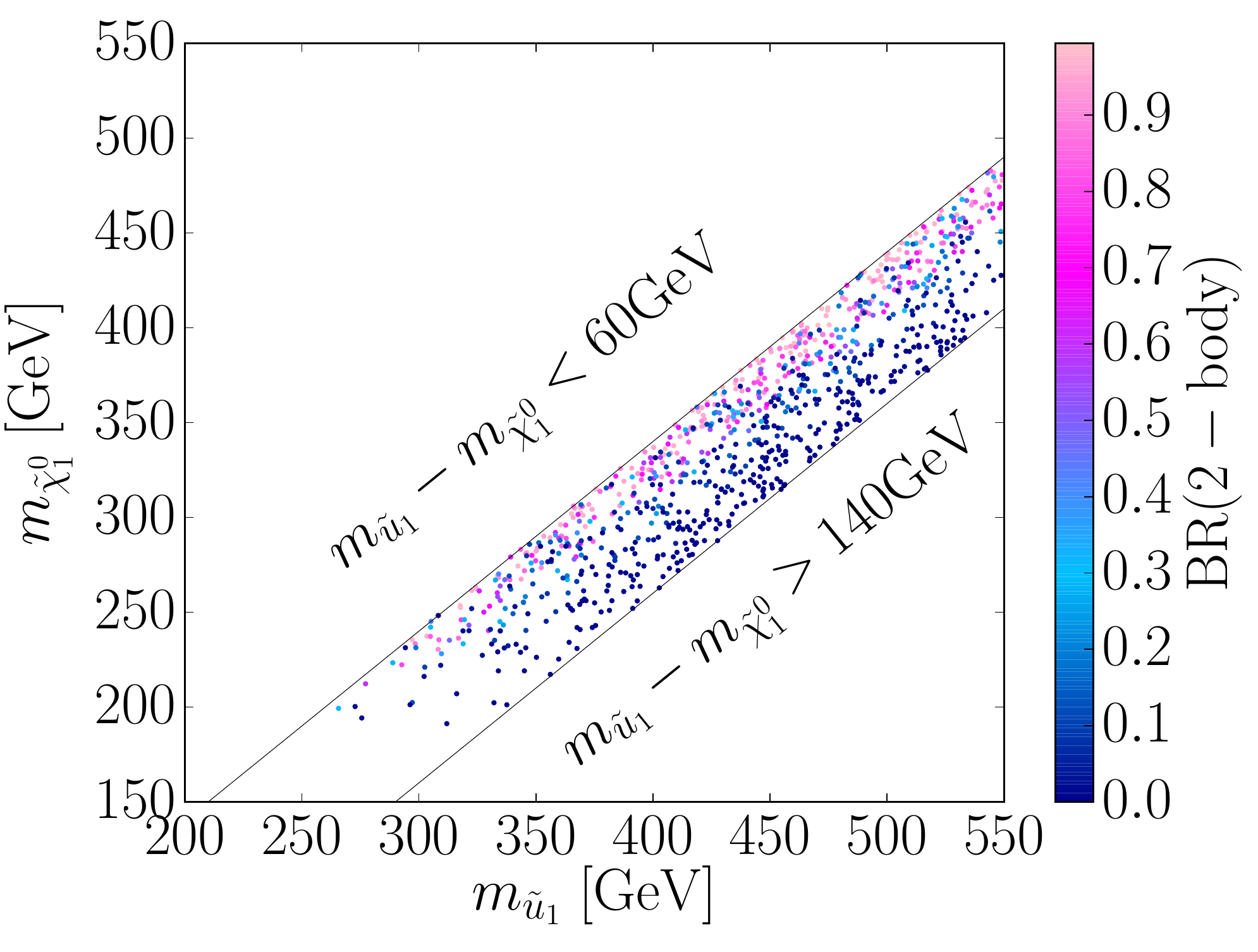}
\end{center}
\vspace{-6mm}\caption{Parameter points of the scan, surviving all
applied constraints, in the $m_{\tilde{u}_1}-m_{\tilde{\chi}_1^0}$
plane. The colour code indicates the corresponding values of the
FV two-body decay branching ratios. The upper (lower) grey line
marks the lower (upper) bound of our investigated mass interval 
below (above) the threshold for the three-body decay.\label{fig:inmassplaneu2}}
\end{figure}

As can be inferred from Fig.~\ref{fig:widths}, the values of the two-body
decay widths are equally distributed along the chosen range of the 
$m_{\tilde{u}_1}-m_{\tilde{\chi}_1^0}$ mass difference. While above
the threshold the three-body decay dominates, close to the threshold
the decay width for the two-body decay, which is shown in red, can be
of similar size as the three- and four-body decay width,
respectively. The branching ratio of the two-body decay is depicted in
Fig.~\ref{fig:2bodybran}. With possible values as large as $\sim 40$\%
at 20~GeV above the threshold, the two-body decay clearly is
competitive with the other decay modes and thereby offers new
discovery perspectives for light stops in this parameter region. In
this region the charm is not soft any more and charm tagging could be
used efficiently, as has been shown in \cite{Aad:2015gna}, 
where a search for pair produced scalar partners of charm quarks was
performed. Such 
large two-body decay widths are achieved in scenarios with relatively
large FV as is the case in the $U(2)$-inspired scenarios 
investigated here. If such a set-up is realised by
nature, Fig.~\ref{fig:2bodybran} shows that it might not be possible
to detect the light stop in the three- and four-body decay mode,
respectively, if the masses of the light stop and the neutralino are 
such that they fall into the threshold regime. Hence, complementary
searches in the two- and the three-, respectively, four-body decay
mode are required in this case. \s

The two-body decay branching ratios for all scenarios of the random
scan that passed the constraints are plotted in
Fig.~\ref{fig:inmassplaneu2} in the
$m_{\tilde{u}_1}$-$m_{\tilde{\chi}_1^0}$ mass plane. The upper and lower grey lines mark the borders of the interval
defined in Eq.~\eqref{eq:numerics1} and the colour code indicates the value of the
two-body decay branching ratio. While for the low mass region the
parameter space is already very constrained such that no valid
parameter points with stop masses lower than $260\ \text{GeV}$ have
been found, the fade out at high values of the stop and neutralino
masses in Fig.~\ref{fig:inmassplaneu2} is due to the limited scan
range of the input parameters of the model. The plot nicely illustrates the
relative importance of the two-body decay in the four- to three-body
transition region and underlines once more the necessity to take this 
decay channel into account in order to allow for a proper analysis of the stop
decays in this mass range. 

\section{Conclusion\label{conclusion}}
In this paper we have analysed decays of the light stop including
the possibility of FV and with the lightest neutralino
being the LSP. We investigated in particular the mass range 
where the three-body decay into $Wb \tilde{\chi}_1^0$ is kinematically
allowed. We provide a proper description of this threshold
region by resorting to the four-body decay into $\tilde{\chi}_1^0 b 
f\bar{f}'$ where the $W$ boson total width has been taken into account in a gauge
invariant way. The resulting decay formula and the three-body decay with
general flavour structure have been included in {\tt
  SUSY-HIT}. We performed a scan over this threshold region where
only the points in 
accordance with the constraints from the LHC Higgs and SUSY data, from
the relic density and $B$ physics measurements as well as from the 
electroweak precision data have been retained. The investigation of
these scenarios revealed that the FV two-body
decay into $c \tilde{\chi}_1^0$ can be comparable to the three-,
respectively, four-body decay and even dominate for some parameter
sets. In order to properly investigate this mass region, the
experiments should therefore also investigate two-body decays with
charm quarks in the final state, in order not to miss the light
stop, which might be the first SUSY particle to be discovered at
the LHC.

\section{Acknowlegdements\label{acknow}}
We are grateful to Ben Allanach and Werner Porod for discussions on
the flavour implementation in their codes. 
M.M.M. would like to thank Filip Moortgat and Michael Spira for discussions on stop
decays. A.W. acknowledges support by the ``Karlsruhe School of Elementary
Particle and Astroparticle Physics: Science and Technology (KSETA)''.

\begin{appendix}
\section{Three-body decay width}
In this appendix we give the analytic formula for the decay width
of the process $\tilde{u}_1\to W d_i \tilde{\chi}_1^0$. We first
define the couplings relevant for the decay width. 
The coupling of an up-type and down-type squark to the $W$ boson, with
the corresponding part of the Lagrangian given by
$\mathcal{L}=g_{W\tilde{u}_s \tilde{d}_t} \tilde{d}_t^{\dagger}
\tilde{u}_s W^-$, is defined as
\begin{equation}
g_{W \tilde{u}_s \tilde{d}_t}=-\frac{g_2}{\sqrt{2}}\sum_{i,j=1}^3
W^{\tilde{u}*}_{s i} V^{CKM*}_{ij} W^{\tilde{d}}_{t j} \, ,
\end{equation}
where $V^{CKM}$ denotes the CKM matrix, $W^{\tilde{u}}$ and
$W^{\tilde{d}}$ the squark mixing matrices in the SCKM basis as
defined in the SLHA2 convention~\cite{Allanach:2008qq} and $g_2$ the
$SU(2)_L$ coupling constant.  
The squark-quark-neutralino couplings to up-type quarks and squarks,
with $\mathcal{L}= \bar{u}_i (g^R_{\chi u_i \tilde{u}_s} P_L+g^L_{\chi
  u_i \tilde{u}_s}P_R) \tilde{u}_s\tilde{\chi}_1^0$, are defined as
\begin{align}
\begin{split}
 g^{L}_{\chi u_i \tilde{u}_s}=&-\sqrt{2}\left(\frac{1}{2} g_2  Z_{12}+ \frac{1}{6} g_1 Z_{11} \right)W^{\tilde{u}*}_{si}\\  &-\frac{g_2}{\sqrt{2}m_W \sin\beta } m_{u_i} Z_{14} W^{\tilde{u}*}_{s i+3}
 \end{split} \\
 \begin{split}
g^R_{\chi u_i \tilde{u}_s}=& \frac{2}{3} \sqrt{2} g_1 Z_{11}^*
W^{\tilde{u}*}_{s i+3}\\ & -\frac{g_2}{\sqrt{2}m_W \sin\beta}
m_{u_i}Z_{14}^* W^{\tilde{u}*}_{si} \, , 
\end{split}
\end{align}
and the ones to down-type squarks and quarks, with $\mathcal{L}= \bar{d}_i (g^R_{\chi d_i \tilde{d}_s} P_L+g^L_{\chi d_i \tilde{d}_s}P_R) \tilde{d}_s\tilde{\chi}_1^0$, as
\begin{align}
\begin{split}
 g^{L}_{\chi d_i \tilde{d}_s}=&\sqrt{2}\left(\frac{1}{2} g_2 Z_{12} - \frac{1}{6} g_1 Z_{11} \right)W^{\tilde{d}*}_{si}\\  &-\frac{g_2}{\sqrt{2}m_W \cos\beta } m_{d_i} Z_{13} W^{\tilde{d}*}_{s i+3}
 \end{split}
 \\
 \begin{split}
g^R_{\chi d_i \tilde{d}_s}=& -\frac{1}{3} \sqrt{2} g_1 Z_{11}^* W^{\tilde{d}*}_{s i+3}\\ & -\frac{g_2}{\sqrt{2}m_W \cos\beta} m_{d_i}Z_{13}^* W^{\tilde{d}*}_{si}\,.
\end{split}
\end{align}
Here, the $W$ boson mass is denoted by $m_W$, $g_1$ is the $U(1)_Y$
coupling constant and $Z$ the neutralino mixing matrix as defined
according to the conventions in Ref.~\cite{Allanach:2008qq}. The angle
$\beta$ is determined through $\tan\beta=v_u/v_d$ where $v_{u,d}$ denote the vacuum expectation values of $H_{u,d}$, respectively. 
The squark-quark-chargino couplings, with $\mathcal{L}=
\bar{d}_i\left(g^L_{\chi^+_l d_i \tilde{u}_s} P_R+ g^R_{\chi^+_l d_i
    \tilde{u}_s}P_L\right)\tilde{u}_s \tilde{\chi}_l^-$, are given by 
\begin{align}
\begin{split}
g^L_{\chi^+_l d_i \tilde{u}_s}  =&\frac{g_2}{\sqrt{2}m_W \sin\beta}V_{l2}\sum_{j=1}^3 m_{u_j}V^{CKM*}_{ji}W^{\tilde{u}*}_{s j+3}\\ &-g_2 V_{l1}\sum_{j=1}^3 W^{\tilde{u}*}_{sj}V^{CKM*}_{ji}
\end{split}
\\
g^R_{\chi^+_l d_i \tilde{u}_s}=&\frac{g_2}{\sqrt{2}m_W\cos\beta}
U_{k2}^* m_{d_i}\sum_{j=1}^3 W^{\tilde{u}*}_{sj}V^{CKM*}_{ji} \, ,
\end{align}
with $U$ and $V$ denoting the chargino mixing matrices as defined in
\cite{Skands:2003cj}. 
The chargino-neutralino-$W$ boson coupling given by
$\mathcal{L}=\bar{\tilde{\chi}}^{-}_l (g^L_{\chi^+_k \chi_1}P_R+
g^R_{\chi^+_k \chi_1}P_L)\tilde{\chi}^0_1 W^-$, is defined as
\begin{align}
 g^L_{\chi^+_k \chi_1}&= g_2\left(Z_{12} V_{k1}^*-\frac{1}{\sqrt{2}}
   Z_{14} V_{k2}^* \right)\\ 
g^R_{\chi^+_k \chi_1}&= g_2\left(Z_{12}^* U_{k1}+\frac{1}{\sqrt{2}}
  Z_{13}^* U_{k2} \right)\,. 
\end{align}
Furthermore we introduce 
\begin{equation}
 \mu_X=\frac{m_X^2}{m_{\tilde{u}_1}^2} \, .
\end{equation}
Note that we will drop the indices for $\tilde{\chi}_1^0$ and denote
the corresponding $\mu$ with $\mu_{\chi}$. By $d_i$ we denote the
final state down-type fermion with flavour index $i$. In
{\tt{SUSY-HIT}} we have set the masses of the first and second generation
quarks to zero, corresponding to $\mu_{d_{1,2}}=0$ and
$\mu_{u_{1,2}}=0$. 
The differential decay width can then be written as
\begin{equation}
\begin{split}
d\Gamma =& \frac{m_{\tilde{u}_1}}{32 (2\pi)^3}\text{Re}\bigg[|\mathcal{M}_{\tilde{d}}|^2+ |\mathcal{M}_{\chi^+}|^2  + | \mathcal{M}_{u}|^2+ \\  & 2\mathcal{M}_{\tilde{d}}\mathcal{M}_{\chi^+}^*+2 \mathcal{M}_{\tilde{d}}\mathcal{M}_u^*+2\mathcal{M}_{\chi^+}\mathcal{M}_u^*\bigg] dx_1 dx_2 \label{eq:diffdec}
\end{split}
\end{equation}
and needs to be integrated over the reduced energies of the final state particles $x_1=2(p_{\tilde{u}_1}\cdot p_{d_i})/m_{\tilde{u}_1}^2$ and
$x_2=2(p_{\tilde{u}_1}\cdot p_{\chi})/m_{\tilde{u}_1}^2$ with $p_X$
denoting the four-momentum of particle $X$. 
The integration limits for the reduced energy $x_1$ are given by
\begin{eqnarray}
 (x_1)_{\textnormal{min}}&=&2 \sqrt{\mu_{d_i}}\\
  (x_1)_{\textnormal{max}}&=&1+\mu_{d_i}-\frac{(m_W+m_{\chi})^2}{m_{\tilde{u}_1}^2} \ .
\end{eqnarray}
For a given value of $x_1$ the range of $x_2$ is determined by
\begin{eqnarray}
 (x_2)_{\textnormal{min}}&=& 1 + \mu_{\chi} - \frac{(E_W + E_{d_i})^2}{m_{\tilde{u}_1}^2}  \\ &+& \frac{\Big(\sqrt{E_W^2-m_W^2}-\sqrt{E_{d_i}^2-m_{d_i}^2}\Big)^2}{m_{\tilde{u}_1}^2} \nonumber \\
 (x_2)_{\textnormal{max}}&=& 1 + \mu_{\chi} - \frac{(E_W + E_{d_i})^2}{m_{\tilde{u}_1}^2}\\ &+& \frac{\Big(\sqrt{E_W^2-m_W^2}+\sqrt{E_{d_i}^2-m_{d_i}^2}\Big)^2}{m_{\tilde{u}_1}^2} \nonumber \ .
 \end{eqnarray}
 Here $E_W$ and $E_{d_i}$ are the energies of the $W$ boson and the down-type quark $d_i$ evaluated in the rest frame of the incoming stop and neutralino,
 \begin{eqnarray}
 E_W&=&\frac{m_{\tilde{u}_1}^2-m_{\tilde{u}_1}^2 x_1+m_{d_i}^2-m_{\chi}^2+m_W^2}{2 \sqrt{m_{\tilde{u}_1}^2-m_{\tilde{u}_1}^2 x_1+m_{d_i}^2}}\\
 E_{d_i}&=&\frac{m_{\tilde{u}_1}^2 x_1 - 2 m_{d_i}^2}{2 \sqrt{m_{\tilde{u}_1}^2-m_{\tilde{u}_1}^2 x_1+m_{d_i}^2}} \ .
 \end{eqnarray} 
The individual contributions to the differential decay width in
Eq.~(\ref{eq:diffdec}) read 
\begin{equation}
\begin{split}
|\mathcal{M}_{\tilde{d}}|^2&= \,8\,\sum_{s,t=1}^{6}  g_{W \tilde{u}_1 \tilde{d}_s}\,g_{W \tilde{u}_1 \tilde{d}_t}^* \\ &\Bigg\{(g_{\chi d_i \tilde{d}_s}^{L}g_{\chi d_i \tilde{d}_t}^{L*}+ 
g_{\chi d_i \tilde{d}_s}^{R}g_{\chi d_i \tilde{d}_t}^{R*}) \\ &
\frac{ y_1 [ 
\mu_W^{-1} (y_2+y_3)^2 -\mu_{\chi} -\mu_{d_i} - 2 y_1]} {(1-x_3 +\mu_W - \mu_{\tilde{d}_s})
(1-x_3 +\mu_W - \mu_{\tilde{d}_t})}\\&+(g_{\chi d_i \tilde{d}_s}^{L}g_{\chi d_i \tilde{d}_t}^{R*}+ 
g_{\chi d_i \tilde{d}_s}^{R}g_{\chi d_i \tilde{d}_t}^{L*})\sqrt{\mu_{d_i}\mu_{\chi}}\\&
\frac{2 y_1+\mu_{\chi}+\mu_{d_i}-\mu_W^{-1}(y_2+y_3)^2}{(1-x_3 +\mu_W - \mu_{\tilde{d}_s})
(1-x_3 +\mu_W - \mu_{\tilde{d}_t})}\Bigg\} \label{gammasbsb}
\end{split}
\end{equation}
\begin{equation}
\begin{split}
|\mathcal{M}_{u}|^2 =& \sum_{j,k=1}^3 \frac{g_2^2\  V^{CKM}_{kn}V^{CKM*}_{jn}}{2(1-x_2 +\mu_{\chi} - \tilde{\mu}_{u_k})(1-x_2 +\mu_{\chi} - \tilde{\mu}_{u_j})} \\&
\bigg\{ 
-2\left(\sqrt{\mu_{u_k}} g^R_{\chi u_k \tilde{u}_1}g^{L*}_{\chi u_j \tilde{u}_1}+(k\leftrightarrow j)\right)
\\ &\sqrt{\mu_{\chi}}\left(\mu_{d_i}+3 y_2 +2 y_2^2 \mu_W^{-1}\right)
\\+ & \ 2 \ g^R_{\chi u_k \tilde{u}_1}g^{R*}_{\chi u_j \tilde{u}_1} \sqrt{\mu_{u_k}\mu_{u_j}} \left(y_1+2 y_2 y_3 \mu_W^{-1}\right)
\\ + & \ 2 \ g^L_{\chi u_k \tilde{u}_1}g^{L*}_{\chi u_j \tilde{u}_1} \left(y_1(\mu_{d_i}-\mu_W+4 y_2)+2 \ y_3 \mu_{d_i} \right. \\ 
+ &  \left.4 \ y_2 y_3 +\mu_W^{-1}(4 y_1 y_2^2-2 y_2 y_3 \mu_{d_i}) \right)
\bigg\}\,,
\end{split}
\end{equation}
with $\tilde{\mu}_{u_3}=\mu_{u_3}+i
\sqrt{\mu_{u_3}}\,\Gamma_t/m_{\tilde{u}_1}$ and $\Gamma_t$ denoting the
top width. For $i=1,2$ the $\tilde{\mu}_{u_i}$ are equal to $\mu_{u_i}$. 
\begin{equation}
\begin{split}
|\mathcal{M}_{\chi^+}|^2 =&  \sum_{k,l=1}^{2} \frac{2}{(1-x_1 - 
\mu_{\chi^+_k}) (1-x_1 - \mu_{\chi^+_l})} \bigg\{  \\
& \left(g^L_{\chi^+_k d_i \tilde{u}_1}g^{L*}_{\chi^+_l d_i \tilde{u}_1}g^L_{\chi^+_k \chi_1}g^{L*}_{\chi^+_l \chi_1}+
(R\leftrightarrow L)
\right) \\ &\bigg[ 4y_3 (y_1+y_2  + \mu_W^{-1} y_1 y_3) + y_1 (\mu_{\chi} -\mu_W) \\ 
&+  2 \mu_{\chi}y_2 (1- \mu_{W}^{-1} y_3) \bigg]  \\ 
+ &\left(g^L_{\chi^+_k d_i \tilde{u}_1}g^{L*}_{\chi^+_l d_i \tilde{u}_1}g^R_{\chi^+_k \chi_1}g^{R*}_{\chi^+_l \chi_1}+(R\leftrightarrow L) \right) \\ 
& \sqrt{\mu_{\chi^+_k} \mu_{\chi^+_l} }\ (y_1+ 2 \mu_W^{-1} y_2 y_3) \\
-& \bigg[ \left(g^L_{\chi^+_k d_i \tilde{u}_1}g^{L*}_{\chi^+_l d_i \tilde{u}_1}g^R_{\chi^+_k \chi_1}g^{L*}_{\chi^+_l \chi_1}+
(R\leftrightarrow L)
\right) \\ &\sqrt{\mu_{\chi^+_k}} + (k\leftrightarrow l)
\bigg] 3 \sqrt{\mu_{\chi}} (y_1+y_2)\\ - & \bigg[\left(g^L_{\chi^+_k d_i \tilde{u}_1}g^{R*}_{\chi^+_l d_i \tilde{u}_1}g^R_{\chi^+_k \chi_1}g^{R*}_{\chi^+_l \chi_1} +
(R\leftrightarrow L)
\right)\\ & \sqrt{\mu_{d_i}\mu_{\chi^+_k}}  +(k\leftrightarrow l)\bigg](\mu_{\chi}+ 2 y_3^2\mu_W^{-1}+ 3 y_3)
\\+& \left(g^L_{\chi^+_k d_i \tilde{u}_1}g^{R*}_{\chi^+_l d_i \tilde{u}_1}g^R_{\chi^+_k \chi_1}g^{L*}_{\chi^+_l \chi_1}+
(R\leftrightarrow L)
\right) \\ & 3 \sqrt{\mu_{d_i}\mu_{\chi}\mu_{\chi^+_l}\mu_{\chi^+_k}}
\\+&
\left(g^L_{\chi^+_k d_i \tilde{u}_1}g^{R*}_{\chi^+_l d_i \tilde{u}_1}g^L_{\chi^+_k \chi_1}g^{R*}_{\chi^+_l \chi_1} +
(R\leftrightarrow L)
\right) \\ & 3\sqrt{\mu_{d_i} \mu_{\chi}}\ (\mu_W+\mu_{\chi}+2
y_3)\bigg\} \, .
\end{split}
\end{equation}
The interference terms read
\begin{equation}
\begin{split}
\mathcal{M}_{\tilde{d}}\mathcal{M}_{\chi^+}^*=& 
\sum_{s=1}^6 \sum_{l=1}^2 \frac{-4 \ g_{W \tilde{u}_1 \tilde{d}_s}}{(1-x_3+\mu_W-\mu_{\tilde{d}_s})
(1-x_1-\mu_{\chi^+_l})} \\ 
&\bigg\{\left(g^{L*}_{\chi^+_l d_i \tilde{u}_1}g^{R*}_{\chi^+_l \chi_1} g^L_{\chi d_i \tilde{d}_s} +(R\leftrightarrow L)\right) \\ 
&\sqrt{\mu_{\chi}\mu_{\chi^+_l}}\ (y_1-y_2 \mu_W^{-1} (y_2+y_3)+\mu_{d_i}) \\ &+ 
\left(g^{L*}_{\chi^+_l d_i \tilde{u}_1}g^{L*}_{\chi^+_l \chi_1} 
g^L_{\chi d_i \tilde{d}_s} +(R\leftrightarrow L)\right)\\&
\bigg[(y_2+y_3)(\mu_\chi y_2- 2 y_1 
y_3) \mu_W^{-1}+y_1 (2 y_1+y_2 \\& - y_3+\mu_{\chi})+\mu_{\chi}y_2-\mu_{d_i}(\mu_{\chi}+y_3)\bigg] \nonumber
\end{split}
\end{equation}
\begin{equation}
\begin{split}
\phantom{\mathcal{M}_{\tilde{d}}\mathcal{M}_{\chi^+}^*= }& 
+(\mu_W^{-1} y_3 (y_2+y_3)-\mu_{\chi}-y_1)\\ &
\bigg[\sqrt{\mu_{d_i} \mu_{\chi}} \left( g^{R*}_{\chi^+_l d_i \tilde{u}_1}g^{R*}_{\chi^+_l \chi_1} 
g^L_{\chi d_i \tilde{d}_s} +(R\leftrightarrow L)\right)+\\&
\sqrt{\mu_{d_i} \mu_{\chi_l^+}} \left( g^{R*}_{\chi^+_l d_i \tilde{u}_1}g^{L*}_{\chi^+_l \chi_1} 
g^L_{\chi d_i \tilde{d}_s} +(R\leftrightarrow L)\right)\bigg]\bigg\}
\end{split}
\end{equation}
\begin{equation}
\begin{split}
\mathcal{M}_{\tilde{d}}\mathcal{M}_u^*=& 
\sum_{s=1}^6 \sum_{j=1}^3\frac{-2 \sqrt{2} \ g_2 \ g_{W \tilde{u}_1 \tilde{d}_s}}{(1-x_3+\mu_{\chi} -\mu_{\tilde{d}_s}) (1-x_2+\mu_{\chi}-\tilde{\mu}_{u_j})} \\ 
& V_{jn}^{CKM*}\Bigg\{ \sqrt{\mu_{u_j} \mu_\chi} \ g^{R*}_{\chi u_j \tilde{u}_1} g^L_{\chi d_i \tilde{d}_s} \, \Bigg[ y_1 + \mu_{d_i}-y_2\\ 
&\mu_W^{-1} (y_2+y_3) \Bigg] +g^{L*}_{\chi u_j \tilde{u}_1} g^L_{\chi d_i \tilde{d}_s} \Bigg[ y_1 y_2  \bigg( 1+ 2\mu_W^{-1} \\ 
&  (y_2+y_3) \bigg)  + \mu_{\chi} y_2 -  y_1 y_3 -2 y_1^2 -y_1 \mu_{d_i}\\ 
& +  \mu_{d_i}(\mu_{\chi}-y_3) + \mu_W^{-1}\mu_{d_i}(-y_2 y_3 -y_3^2) \Bigg] \\ 
+ & \ g^{R*}_{\chi u_j \tilde{u}_1} g^R_{\chi d_i \tilde{d}_s} \sqrt{\mu_{d_i}\mu_{u_j}} \bigg[- y_1-\mu_{\chi} \\
+ &\ \mu_W^{-1} y_3 (y_2+y_3) \bigg] + g^{L*}_{\chi u_j \tilde{u}_1} g^{R}_{\chi d_i \tilde{d}_s} \sqrt{\mu_{d_i}\mu_{\chi}} \\ 
&\bigg[ y_1+\mu_{d_i} -\mu_W^{-1} y_2 (y_2+y_3) \bigg] \Bigg\} 
\end{split}
\end{equation}
\begin{equation}
\begin{split}
\mathcal{M}_{\chi^+}\mathcal{M}_u^*=& 
\sum_{k=1}^2  \sum_{j=1}^3\frac{g_2}{\sqrt{2}(1-x_1 -\mu_{\chi_k^+})(1-x_2+\mu_{\chi}-\tilde{\mu}_{u_j})} \\ 
& V_{jn}^{CKM*} \Bigg\{ g^L_{\chi^+_k d_i \tilde{u}_1}g^R_{\chi^+_k \chi_1} g^{L*}_{\chi u_j \tilde{u}_1} \sqrt{ \mu_{\chi} \mu_{\chi_k^+} } \\ 
& (-6 y_2- 4  \mu_W^{-1} y_2^2 - 2 \mu_{d_i}) + 2\ g^L_{\chi^+_k d_i \tilde{u}_1}g^L_{\chi^+_k \chi_1} g^{L*}_{\chi u_j \tilde{u}_1} \\ 
& \Bigg[ y_1(2 y_3+ 2 y_2 +  4 y_1-\mu_W)+  y_2 (4 y_3+\mu_{\chi})\\
&- 2 y_2 (2 y_1 y_3-\mu_{\chi} y_2)\mu_W^{-1}+ 2 \mu_{d_i}\mu_W^{-1}y_3^2\\
&-\mu_{d_i} \mu_{\chi} + \mu_{d_i} y_3 \Bigg]\\
- &\ 6\ \sqrt{ \mu_{u_j} \mu_{\chi} }\ g^L_{\chi^+_k d_i \tilde{u}_1}g^L_{\chi^+_k \chi_1} g^{R*}_{\chi u_j \tilde{u}_1} (y_1+y_2)\\ 
+ & \sqrt{ \mu_{u_j} \mu_{\chi^+_k} }\ g^L_{\chi^+_k d_i \tilde{u}_1}g^R_{\chi^+_k \chi_1} g^{R*}_{\chi u_j \tilde{u}_1} (2 y_1 + 4 \mu_W^{-1} y_2 y_3) \nonumber
\end{split}
\end{equation}
\begin{equation}
\begin{split}
\phantom{\mathcal{M}_{\chi^+}\mathcal{M}_u^*=}
+ &\ g^R_{\chi^+_k d_i \tilde{u}_1}g^R_{\chi^+_k \chi_1} g^{R*}_{\chi u_j \tilde{u}_1} \sqrt{\mu_{d_i}\mu_{u_j}}\\ 
& (-6 y_3 - 4 y_3^2 \mu_W^{-1} - 2\mu_{\chi})\\ 
+ & \ 6 g^R_{\chi^+_k d_i \tilde{u}_1}g^L_{\chi^+_k \chi_1} g^{R*}_{\chi u_j \tilde{u}_1}\sqrt{\mu_{d_i}\mu_{u_j}\mu_{\chi}\mu_{\chi_l^+}} \\ 
- & \ 6  g^R_{\chi^+_k d_i \tilde{u}_1}g^L_{\chi^+_k \chi_1} g^{L*}_{\chi u_j \tilde{u}_1} \sqrt{\mu_{d_i}\mu_{\chi_l^+}}\ (y_1+ y_3)\\ 
+  & \ g^R_{\chi^+_k d_i \tilde{u}_1}g^R_{\chi^+_k \chi_1} g^{L*}_{\chi u_j \tilde{u}_1} \sqrt{\mu_{d_i}\mu_{\chi}}  \bigg[6 \mu_W+6 y_3\\ 
+ & \ 6 y_2+2 y_1+4 \mu_W^{-1} y_2 y_3) \bigg] \Bigg\}\,.\label{gammatchi}
\end{split}
\end{equation}
 In Eqs.~\eqref{gammasbsb}--\eqref{gammatchi} we have used
\begin{align}
x_3 &= 2 - x_1 -x_2 \\
y_1 & = \frac{1}{2}(1+\mu_W-\mu_{\chi}-\mu_{d_i}-x_3)\\
y_2 & = \frac{1}{2}(1-\mu_W+\mu_{\chi}-\mu_{d_i}-x_2)\\
y_3 & = \frac{1}{2}(1-\mu_W-\mu_{\chi}+\mu_{d_i}-x_1)\;.
\end{align}
The notation $(R\leftrightarrow L)$ means that the respective term is obtained from the previous one by replacing $R \leftrightarrow L$ in the couplings, whereas $(k\leftrightarrow l)$ and $(k\leftrightarrow j)$ means that the term is obtained by interchanging indices $k$ and $l$ or $k$ and $j$, respectively.
\end{appendix}
\newpage

\bibliographystyle{utphys}

\end{document}